\documentclass[useAMS,usenatbib,usegraphicx]{mn2e}
\usepackage{amssymb,amsmath,bm} 
\usepackage{natbib}
\usepackage{graphics}
\graphicspath{{./fig/}{./png/}}

\usepackage{color}

\newcommand{\Eq}[1]{Eq.~(\ref{#1})}
\newcommand{\EQ}{\begin{equation}}
\newcommand{\EE}{\end{equation}}
\newcommand{\EQA}{\begin{eqnarray}}
\newcommand{\EEA}{\end{eqnarray}}
\newcommand{\brac}[1]{\langle #1 \rangle}
\newcommand{\pd}{\partial}

\newcommand{\DIV}{\nab \cdot }

\newcommand{\mean}[1]{\overline{#1}}

\newcommand{\Sec}[1]{Section~\ref{#1}}
\newcommand{\Fig}[1]{Figure~\ref{#1}}
\newcommand{\Tab}[1]{Table~\ref{#1}}
\newcommand{\kf}{k_{\rm f}}
\newcommand{\cs}{c_{\rm s}}
\newcommand{\cst}{c_{\rm s}^2}

\newcommand{\etat}{\eta_{\rm t}}
\newcommand{\etatz}{\eta_{\rm t0}}

\newcommand{\urms}{u_{\rm rms}}

\newcommand{\brms}{b_{\rm rms}}

\newcommand{\Beq}{B_{\rm eq}}
\newcommand{\Ma}{\rm Ma}
\newcommand{\ii}{\rm i}

\newcommand{\meanh}{\overline{h}}
\newcommand{\meanhm}{\overline{h}_{\rm m}}
\newcommand{\meanhf}{\overline{h}_{\rm f}}
\newcommand{\kef}{k_{\rm f}}
\newcommand{\keff}{k_{\rm eff}}

\newcommand{\St}{{\rm St}}

\newcommand{\Shw}{{\it S}_{\it W}}

\newcommand{\Rm}{{{\rm Rm}}}
\newcommand{\Rmast}{{{\rm Rm}_{\ast}}}

\newcommand{\nab}{\mbox{\boldmath $\nabla$} {}}
\newcommand{\meanFFFF}{\overline{\mbox{\boldmath ${\cal F}$}}{}}{}
\newcommand{\meanFFFFm}{\overline{\mbox{\boldmath ${\cal F}$}}_{\rm m}{}}{}
\newcommand{\meanFFFFf}{\overline{\mbox{\boldmath ${\cal F}$}}_{\rm f}{}}{}
\newcommand{\meanFFFfz}{\overline{\cal F}_{\rm f{z}}}{}
\newcommand{\meanFFFdz}{\overline{\cal F}_{z}^{\rm diff}}{}
\newcommand{\meanFFF}{\overline{\cal F}}{}
\newcommand{\memf}{\overline{\mbox{\boldmath ${\cal E}$}}{}}{}
\def\onethird{{\textstyle{1\over3}}}
\def\onehalf{{\textstyle{1\over2}}}

\newcommand{\EEE}{\bm{E}}
\newcommand{\AAA}{\bm{A}}
\newcommand{\aaa}{\bm{a}}
\newcommand{\bb}{\bm{b}}
\newcommand{\uu}{\bm{u}}
\newcommand{\ff}{\bm{f}}
\newcommand{\jj}{\bm{j}}
\newcommand{\BB}{\bm{B}}
\newcommand{\UU}{\bm{U}}
\newcommand{\JJ}{\bm{J}}
\newcommand{\zz}{\bm{z}}
\newcommand{\dd}{{\rm{d}}}
\newcommand{\DD}{{\rm{D}}}
\newcommand{\meanAA}{\overline{\bm{A}}}
\newcommand{\meanBB}{\overline{\bm{B}}}
\newcommand{\meanUU}{\overline{\bm{U}}}
\newcommand{\meanB}{\overline{\bm{B}}}
\newcommand{\meanU}{\overline{\bm{U}}}
\newcommand{\meanrho}{\overline{\rho}}
\newcommand{\meanBrms}{\overline{\bm{B}}_{\rm rms}}

\title
[Advective magnetic helicity flux]
{Turbulent dynamos with advective magnetic helicity flux}

\author[F. Del Sordo, G. Guerrero  and A. Brandenburg]
{F. Del Sordo$^{1,2}$\thanks{E-mail: 
fabio@nordita.org (FDS)} and
G. Guerrero$^{3,1}$ and
A. Brandenburg$^{1,2}$
\\
$^{1}$Nordita, KTH Royal Institute of Technology and Stockholm University,
Roslagstullsbacken 23, SE 10691 Stockholm Sweden \\
$^{2}$Department of Astronomy, AlbaNova University Center, Stockholm
University, SE 10691 Stockholm, Sweden  \\
$^{3}$Solar Physics, HEPL, Stanford University, Stanford, CA, 
94305-4085, USA
}
\begin{document}


\pagerange{\pageref{firstpage}--\pageref{lastpage}} \pubyear{2002}

\maketitle

\label{firstpage}

\begin{abstract}
Many astrophysical bodies harbor magnetic fields that are 
thought to be sustained by a dynamo process.
However, it has been argued that the production of large-scale magnetic
fields by mean-field dynamo action is strongly suppressed at large
magnetic Reynolds numbers owing to
the conservation of magnetic helicity.
This phenomenon is known as {\it catastrophic quenching}.
Advection of magnetic fields by stellar and galactic winds
toward the outer boundaries and away
from the dynamo is expected to alleviate such quenching.
Here we explore the relative roles played by advective and 
turbulent--diffusive fluxes of magnetic helicity in the dynamo.
In particular, we study how the dynamo is affected by advection. 
We do this by performing direct numerical simulations of a
turbulent dynamo of $\alpha^2$ type driven by forced turbulence
in a Cartesian domain in the presence of
a flow away from the equator where helicity changes sign.
Our results indicate that in the presence of advection, the dynamo, 
otherwise stationary, becomes oscillatory.
We confirm an earlier result for turbulent--diffusive magnetic helicity fluxes
that for small magnetic Reynolds numbers ($\Rm\lesssim 100...200$, based
on the wavenumber of the energy-carrying eddies) the magnetic helicity flux
scales less strongly with magnetic Reynolds number ($\Rm^{-1/2}$) than the
term describing magnetic helicity destruction by resistivity ($\Rm^{-1}$).
Our new results now suggest that for larger $\Rm$ the former becomes
approximately independent of $\Rm$, while the latter falls off more slowly.
We show for the first time that both for weak and stronger winds, the magnetic 
helicity flux term becomes comparable to the resistive term for
$\Rm\gtrsim 1000$, which is necessary for alleviating catastrophic quenching.
\end{abstract}


\section{Introduction}
A theoretical framework for explaining
the large-scale magnetic fields observed in many
astrophysical bodies is mean-field dynamo theory. 
Its basic idea is that the inductive effects of turbulent motions are 
able to amplify a weak magnetic field and maintain it on timescales
longer than the magnetic diffusion time \citep{moffatt_78}.
Gradients in the large-scale velocity field, like shear motions,
can also contribute significantly to the amplification of the magnetic field.  
In mean-field dynamo theory the contribution of the turbulent
scales is parameterized through the electromotive force
which depends on the large-scale magnetic field 
as well as its derivatives \citep{KR80}.
The coefficients in front of the magnetic field and its derivatives are
called turbulent transport coefficients.
They can describe either turbulent--diffusive
(with turbulent diffusion $\propto\etat$)
or non-diffusive (e.g., the $\alpha$ effect or turbulent pumping) effects.

Under some approximations (e.g., in the low conductivity limit for small
magnetic Reynolds number, $\Rm\le1$, or in the high conductivity limit
for short correlation times, i.e.,
small Strouhal number, $\St\le1$), theories like the first
order smoothing approximation are able to predict the functional 
form of the expressions and the correct values of the coefficients.
Within their limits of validity,
these results present a remarkably good agreement 
with the computation of the transport coefficients through direct 
numerical simulations (DNS); see, e.g., \cite{SBS08}.
However, not enough is known about the functional form of
these coefficients at large values of $\Rm$ (i.e., small values
of the microphysical magnetic diffusivity) and about the
saturation process when the magnetic field becomes
dynamically important.
Understanding the behavior of the dynamo in these regimes
has remained an important problem for several decades.
Although many recent works have contributed to
understanding dynamo saturation at large magnetic Reynolds numbers,
more work is still necessary 
to have a complete picture of the dynamo excitation and saturation 
mechanisms.

Among the turbulent transport coefficients the $\alpha$ effect is
particularly important, because it allows a closed dynamo loop
for regenerating both poloidal and toroidal magnetic fields.
It has been suspected,
however, that in closed or triply periodic domains the $\alpha$ effect
can be strongly suppressed at higher magnetic Reynolds numbers 
and might scale like $\alpha\propto\Rm^{-1}$ \citep{VC92,CH96}.
An explanation for this was proposed by \cite{GD94}, who used 
the $\alpha$ effect
derived by \cite{PFL76}, which has, in addition to the kinetic helicity
density, a contribution proportional to the current helicity density.
It is this quantity which builds up as the dynamo saturates.

This is a consequence of magnetic helicity conservation and
can be explained as follows:
the large-scale magnetic field generated by the $\alpha$ effect is 
helical, but in order to satisfy the conservation of total 
magnetic helicity, a small-scale field with equally strong magnetic helicity
of opposite sign must be generated in the 
system. The small-scale magnetic helicity is responsible for 
the creation of a
magnetic $\alpha$ effect ($\alpha_{\rm M}$) which contributes 
with opposite sign to the kinetic $\alpha$.
This basic idea led \cite{KR82} to propose the dynamical quenching
model at a time well before simulations saw any indications of
catastrophic quenching.
Even nowadays the issue is quite unclear when it comes to making predictions
about the high-$\Rm$ regime.
The final amplitude that the magnetic $\alpha$ effect acquired depends on the
geometry of the system and on the value of the magnetic Reynolds number.
For highly turbulent astrophysical objects
with high $\Rm$ like the Sun or the Galaxy, $\alpha_{\rm M}$ could
attain higher amplitudes, decreasing then the dynamo efficiency. However,
the dynamics of $\alpha_{\rm M}$ also depends on the ability
of the system to get rid of the small-scale magnetic helicity 
responsible for its creation.
In a closed or triply periodic homogeneous domain, magnetic helicity
annihilation depends just on the microscopic magnetic diffusivity.
This is a very slow process given the scales and diffusivity values
under consideration.
However, an obvious solution to this catastrophic ($\Rm$-dependent) 
quenching is to allow the system to get rid of helical small-scale magnetic
fields.

In real astrophysical systems, this processes is generally expected to
happen in a number of different ways.
Among the various mechanisms for removing magnetic helicity from the
system we focus here on the role played by the turbulent--diffusive magnetic 
helicity flux and by the presence of advective flows or winds.
The role of these magnetic
helicity fluxes has been tested in the context of mean-field
dynamo models through a dynamical equation for the magnetic
$\alpha$-effect \citep{KMRS00,BS05c,SSSB06,SSS07,BCC09,GCB10,CGB11}.
These models have demonstrated the importance of magnetic helicity
fluxes in solving the catastrophic quenching problem.

Verifying the validity of these results in DNS
is more complicated since obtaining higher 
$\Rm$ in the numerical models requires high resolution and large
computational resources.
Various attempts have, however,
succeeded in demonstrating the role of magnetic helicity conservation
in the saturation of the dynamo.
For instance, \cite{B01} 
studied the saturation in triply periodic helically forced dynamos
of $\alpha^2$ type.
The role of open magnetic boundary conditions  for
convective dynamos has been studied in \cite{KKB08,KKB09,KKB10b}.
Furthermore, using forced turbulence, \cite{Mitra+etal_10} (hereafter MCCTB)
have verified the existence of turbulent--diffusive
magnetic helicity fluxes in $\alpha^2$ dynamo models in the presence
of an equator and \cite{HB10} (hereafter HB) did the same for a
dynamo region embedded inside a highly conducting halo which provided a more 
realistic boundary condition.
In both cases it was found that a fit to a Fickian diffusion law can
account for this flux and that the diffusivity value is comparable to
or below the value of the turbulent magnetic diffusivity $\etat$.
The resulting Fickian diffusion coefficient was found to be
approximately independent of $\Rm$.
By considering a statistically steady state, and noting that the local
value of the magnetic helicity density was also statistically steady,
their result became then also independent
of the gauge chosen to define the magnetic vector potential.

In addition, shear flows have been argued to be 
effective in alleviating catastrophic quenching \citep{VC01}
and allowing significant saturation levels of the dynamo \citep{KKB08},
although it appears now plausible that their result could also be
explained through a change in the excitation conditions of the dynamo.
Indeed, recent DNS have failed to demonstrate the presence
of the Vishniac-Cho flux \citep{HB11}.
Yet another possibility is the advective magnetic helicity flux.
In the context of the galactic dynamo, alleviation of catastrophic 
quenching thanks to a wind
has been studied in mean-field models by \cite{SSSB06} 
and \cite{SSS07}.
\cite{MMTB11} studied the role of a wind
in solar mean-field dynamo models. 
The models studied in the present paper allow us to compare with
their results and to determine the importance of magnetic helicity fluxes
in the dynamical evolution of the magnetic $\alpha$-effect.
To our understanding the study of advective 
fluxes in DNS of a dynamo is an outstanding problem.
With this paper we intend to close this gap.    

We perform DNS leading to $\alpha^2$-type dynamo action in a domain
with kinetic helicity of opposite signs on both sides of the equator.
We use a relaxation term to include a 
large-scale flow that advects the large-scale magnetic field.
Furthermore, we consider periodic boundary conditions in the
horizontal directions, zero-gradient conditions for the velocity
and vertical field conditions for the magnetic field.
In this way we allow for the removal of magnetic helicity through
advection.  For the sake of simplicity and to study these 
effects separately in a clear way,
we do not include large-scale shear.
Nevertheless the results presented here should also be 
applicable in the context of the galactic dynamo and, in principle, also
to the solar dynamo, where large-scale winds have been shown in
mean-field models to play a role in carrying magnetic helicity 
outside its bounds \citep{MMTB11}.

This paper is organized as follows. In Sect. \ref{sect:model} we
describe the physical model considered here and present the equations 
governing its evolution.  In Sect. \ref{sect:results} we present 
the results of the simulations. First we describe the properties of
the solutions without the wind. Next, we explore the effects that
the wind has on the characteristics of the dynamo solution. Finally, we
determine the magnetic helicity fluxes present in the model and 
verify their balance with the production terms to prevent the 
quenching of what corresponds to the $\alpha$ effect in the related
mean-field description.  We conclude and summarize the
results in Sect. \ref{sect:concl}.
 
\section{The model}
\label{sect:model}

\subsection{Governing equations}

We use the \textsc{Pencil Code}%
\footnote{\texttt{http://pencil-code.googlecode.com/}}
to solve the following set of compressible hydromagnetic equations in 
an isothermal layer:
\begin{equation}
\frac{\pd \bm A}{\pd t} = \bm{U}\times\bm{B}-\mu_0\eta {\bm J},\label{equ:AA}
\end{equation}
\begin{equation}\label{equ:UU}
 \frac{\DD \bm{U}}{\DD t} = -\cst \nab \ln \rho +
 \frac{1}{\rho} \bm{J} \times {\bm B} + \frac{1}{\rho} \bm{\nabla}
 \cdot 2 \nu \rho \mbox{\boldmath ${\sf S}$}
 +\bm{f}_{\rm w} + \bm{f}, 
\end{equation}
\begin{equation}\label{equ:rho}
\frac{\DD \ln \rho}{\DD t} = -\nab\cdot{\bm U} + q_\rho,
\end{equation}
where $\DD/\DD t = \pd/\pd t + \bm{U} \cdot
\bm{\nabla}$ is the advective derivative,
$\bm{A}$ is the magnetic vector potential, $\bm{B} =
\bm{\nabla} \times \bm{A}$ is the magnetic field, $\bm{J}
=\bm{\nabla} \times \bm{B}/\mu_0$ is the current density, $\mu_0$ is
the magnetic permeability, $\eta$ and $\nu$ 
are magnetic diffusivity and kinematic viscosity, respectively,
$\cs=\mbox{const}$ is the sound speed,
$\bm{U}$ is the velocity, $\rho$ is the density,
$\mbox{\boldmath ${\sf S}$}$ is the rate of strain tensor given by
\begin{equation}
{\sf S}_{ij} = \onehalf (U_{i,j}+U_{j,i}) - \onethird \delta_{ij} \DIV \bm{U},
\end{equation}
where the commas denote derivatives, 
$\bm{f}_{\rm w}$ provides a forcing for the wind
(defined below in \Sec{sect:wind}),
$q_\rho$ is a source term in \Eq{equ:rho} needed to replenish the
resulting mass loss,
$\bm{f}$ is a time-dependent random $\delta$-correlated
forcing function of the form
\begin{equation}
\bm{f} = \bm{f}(\bm{x},t;\sigma(z)), \label{equ:hf}
\end{equation}
where $\sigma$ is related to its local helicity density,
\begin{equation}
\brac{\ff\cdot\nab\times\ff}/\brac{\kf\ff^2}=2\sigma/(1+\sigma^2),
\end{equation}
and is chosen to vary
like $\sigma(z) = \sin(2\pi z/L_z)$ with a sign change across the equator
at $z=0$.
This forcing drives turbulence in a band of wavenumbers around $\kf$.
The modulation $\sigma(z)$ of this forcing is similar to that used by 
\cite{WBM11} to simulate
a sign change of helicity in forced turbulence in a spherical wedge.

We consider a computational domain of size $L_x\times L_y\times L_z$,
with quadratic horizontal extent, $L_x=L_y$, using periodic boundary conditions
and a vertical extent that is twice as big, $L_z=2L_x$, with
$|z|\leq L_z/2$ (i.e., $-L_z/2\leq z\leq L_z/2$) and an equator at $z=0$.
Our boundary conditions are
\begin{equation}
U_{x,z}=U_{y,z}=U_z-\mean{U}_{\rm w}=A_{x,z}=A_{y,z}=A_z=0
\end{equation}
on the top and bottom boundaries at $z=\pm L_z/2\equiv \pm z_{\rm top}$.
$\mean{U}_{\rm w}$ is the wind profile, defined below in \Sec{sect:wind}.
The lowest horizontal wavenumber in the domain is $k_1=2\pi/L_x$.
In the following, we use $k_1$ as our inverse length unit,
so $|k_1 z|\leq 2\pi$.
To eliminate boundary effects, we restrict most of the analysis
to a diagnostic layer, $|z|\leq L_*$ with $k_1 L_*=3$.
For all our runs we choose $\kf/k_1=4$, which is a compromise between it
being large enough to allow a large-scale magnetic field to be generated
and yet small enough to achieve sufficiently large values of $\Rm$.

We set $\cs$ to unity in the code, so our dimensionless time
is in units of the sound travel time, $(\cs k_1)^{-1}$.
However, the relevant physics is not governed by compressibility effects,
so it is more natural to quote time in turnover times, i.e., we quote 
instead the value of $t\urms\kf$.
In most of the cases reported below, the turbulent Mach number,
$\Ma=\urms/\cs$, is around 0.1.
Likewise, in the code $\nu$ and $\eta$ are given in units of $\cs/k_1$,
but it is physically more meaningful to quote corresponding Reynolds
numbers instead.
Our resolution is increased with increasing values of $\Rm$, so the
largest resolution used in this paper is $1024\times1024\times2048$
meshpoints.
We return to this issue at the end of the paper.

\subsection{Generating the wind}
\label{sect:wind}

In our model, the advective term from the wind is given by the forcing
function in \Eq{eq:wind},
\begin{equation}
\bm{f}_{\rm w}=-\frac{1}{\tau_{\rm w}}
\left[\meanUU-\meanUU_{\rm w}(z)\right],
\label{eq:wind}
\end{equation}
where $\meanUU$ is the horizontally averaged velocity field, and
\begin{equation}\label{eq:windprof}
\meanUU_{\rm w}(z)=U_0\frac{\zz}{z_{\rm top}}
\end{equation}
is the wind profile that increases linearly
toward the $z$ boundaries.
The wind profile can be modified by the turbulence and the magnetic field,
but the original outflow profile tends to be restored
on a timescale $\tau_{\rm w}$.
The presence of a wind leads to mass loss across the vertical boundaries
with a mass loss rate that depends on $U_0$. 

Stellar winds are the main agents of mass loss in stars.
In a galactic environment it is possible to observe galactic winds
as well as galactic fountains.
These mechanisms can be driven by the explosions of supernovae
in the galactic disc.
In this case a direct estimate of the mass loss rate is more complicated, 
given that it is expected to be very small.
However, to have stationary conditions, we keep the mass
in the domain constant using the source term $q_\rho$ in \Eq{equ:rho}.
This source term tends to be restored the density at each
spatial point in the domain to its initial value $\rho_0$
on a timescale $\tau_{\rm s}=\tau_{\rm w}$.
Thus, analogously to \Eq{eq:wind}, we write
$q_\rho=-\tau_s^{-1}(\ln\meanrho-\ln\meanrho_0)$.

We study the dependence of our model on the dimensionless wind speed
and the magnetic Reynolds number of the turbulence.
These are defined as
\begin{equation}
\Shw=\frac{\nab\cdot\mean{U}_{\rm w}}{\urms\kf},\quad
\Rm=\frac{\urms}{\eta\kf}.
\end{equation}
In all cases, we use a magnetic Prandtl number of unity, i.e., $\nu/\eta=1$.

\subsection{Magnetic helicity fluxes}

In our model we expect two different kinds of magnetic helicity fluxes:
those caused by the wind, i.e.\ \emph{advective} magnetic helicity fluxes,
and those due to turbulence in the presence of a mean gradient of the magnetic
helicity density, i.e.\ \emph{turbulent--diffusive} magnetic helicity fluxes.
To assess their importance in the magnetic helicity budget,
we now consider the magnetic helicity 
equation in the Weyl gauge which is used in \Eq{equ:AA}, i.e.,
\begin{equation}
{\partial\over\partial t}\overline{\AAA\cdot\BB}=
-2\eta\mu_0\overline{\JJ\cdot\BB}-\nab\cdot\meanFFFF,
\end{equation}
where overbars denote averages over $x$ and $y$
and $\meanFFFF=\overline{\EEE\times\AAA}$ is the total magnetic helicity
flux, with $\EEE=\eta\mu_0\JJ-\UU\times\BB$ being the electric field
in the lab frame.
This equation is evidently gauge-dependent; see for instance \cite{CHBK11}.
In particular, since $\overline{\AAA\cdot\BB}$ is not a physical quantity,
it could drift -- even in the steady state; see Fig.~2 of \cite{BDS02} for
an example.
However, {\em if} $\overline{\AAA\cdot\BB}$ is constant in a particular gauge,
then we have
\begin{equation}
\nab\cdot\meanFFFF=
-2\eta\mu_0\overline{\JJ\cdot\BB},
\end{equation}
where now $\nab\cdot\meanFFFF$ must be gauge-independent, because
$\JJ$ and $\BB$ are gauge-invariant.
This argument was invoked by MCCTB and HB to
determine turbulent--diffusive contributions to the magnetic helicity flux.

In the present work, we are interested in two contributions
to $\meanh=\overline{\AAA\cdot\BB}$,
one from the mean fields, $\meanhm=\meanAA\cdot\meanBB$,
and one from the fluctuating fields, $\meanhf=\overline{\aaa\cdot\bb}$.
Their sum gives the total mean magnetic helicity density,
i.e., $\meanh=\meanhm+\meanhf$.
Note, however, that only $\meanhf$ is the component directly relevant for the 
study of catastrophic quenching, because it is approximately
proportional to the current helicity
density, $\overline{\jj\cdot\bb}$, which in turn determines the
magnetic contribution to the $\alpha$ effect.
[The approximate proportionality of magnetic and current helicities
is non-trivial and will need to be re-assessed below; see also Fig.~3 of
MCCTB and Table~2 of HB for earlier examples.]

The evolution equation for $\meanhf$ is
\begin{equation}
{\partial\meanhf\over\partial t}=
-2\memf\cdot\meanBB-2\eta\mu_0\,\overline{\jj\cdot\bb}
-\nab\cdot\meanFFFFf,
\end{equation}
where, as mentioned above, we allow two contributions to the flux of
magnetic helicity from the fluctuating field
$\meanFFFFf$: an advective flux due to the wind,
$\meanFFFFf^{\rm w}=\meanhf\meanU_{\rm w}$,
and a turbulent--diffusive flux due to turbulence,
modelled here by a Fickian diffusion term down the gradient
of $\meanhf$, i.e., $\meanFFFFf^{\rm diff}=-\kappa_h\nab\meanhf$.
Here, $\memf=\overline {\uu \times \bb} $ is the electromotive force
of the fluctuating field.

In the steady state, and if $\meanhf$ is then also constant (which is not
guaranteed to be the case because $\meanhf$ is {\it a priori} gauge-dependent),
we have
\begin{equation}
\nab\cdot\meanFFFFf=-2\memf\cdot\meanBB-2\eta\mu_0\,\overline{\jj\cdot\bb}.
\label{balance}
\end{equation}
Again, although $\nab\cdot\meanFFFFf$ is in principle gauge-dependent, 
it can now be
determined by measuring $\memf\cdot\meanBB$ and $\overline{\jj\cdot\bb}$,
that are manifestly gauge-independent quantities. 
This means that $\nab\cdot\meanFFFFf$ must be gauge-independent as well.
We assume that $\meanFFFFf$ has a component only in the vertical direction.
We can therefore obtain its $z$ dependence through integration via
\begin{equation}
\meanFFFfz=\int_0^z\nab\cdot\meanFFFFf\,\dd z'.
\end{equation}
The assumption of only a $z$ component of $\meanFFFFf$
would break down in the presence of shear, where
cross-stream fluxes with finite divergence are possible; see \cite{HB11}.

For the discussion of our results presented below, let us contrast
our present simulations with those of MCCTB.
In their case, the outer boundary condition at $z=\pm L_z/2$
was a perfect conductor (P.C.) one and the most easily excited mode
was antisymmetric about the midplane with dynamo waves propagating
toward the equator.
This antisymmetry results in permitting a flux of magnetic
helicity through the equatorial plane and in this sense
has the same effect as the vertical field (V.F.) boundary condition.
This, together with the fact that the magnetic helicity density is
antisymmetric about the equator, is the reason why in their case the
turbulent--diffusive flux can play a measurable role.
However, because $\meanFFF_{{\rm f}z}$ has vanishing vertical derivative
at the equator, the $\nab\cdot\meanFFFFf$ vanishes there.
This is different in the model of HB, in which the helicity is arranged
to be symmetric about the midplane,
which is therefore not an equator in the usual sense.
Here the field is symmetric about the midplane, corresponding thus
to a P.C.\ condition, and thus $\nab\cdot\meanFFFFf\neq0$.
The boundary conditions and their properties are summarized in \Tab{table:bc}
for MCCTB and HB and compared with those used in the present work.

\begin{table}\caption{
\label{table:bc}
Comparison of boundary conditions and other properties of the simulations
of MCCTB and the present work.
}\centerline{\begin{tabular}{l|lll}
  &  MCCTB & HB & present work \\
\hline
boundary   & P.C. & halo & V.F. \\
           & $\meanFFF_{{\rm f}z}=0$ 
           & $\meanFFF_{{\rm f}z}\neq0$ 
           & $\meanFFF_{{\rm f}z}\neq0$ \\
           & $\nab\cdot\meanFFFFf\neq0$ 
           & $\nab\cdot\meanFFFFf=0$ 
           & $\nab\cdot\meanFFFFf=0$ \\
\hline
equator/   & antisymmetry & symmetry & symmetry \\
midplane   & (like V.F.) & (like P.C.) & (like P.C.) \\
           & $\meanFFF_{{\rm f}z}\neq0$
           & $\meanFFF_{{\rm f}z}=0$
           & $\meanFFF_{{\rm f}z}=0$ \\
           & $\nab\cdot\meanFFFFf=0$
           & $\nab\cdot\meanFFFFf\neq0$
           & $\nab\cdot\meanFFFFf=0$ \\
\label{Symmetry}\end{tabular}}\end{table}

Unlike MCCTB, in the present work the V.F.\ condition is
applied on the outer boundaries, in which case the most easily excited
mode is symmetric about the equator with dynamo waves travelling away
from the midplane.
This is similar to a P.C.\ condition at the midplane,
for which the magnetic helicity flux vanishes.
However, because $\meanhf$ is antisymmetric about the equator,
it must have a turning point there, so its second derivative
vanishes and $\nab\cdot\meanFFFFf=0$.
The present model does not have shear, but the nature of the dominant
mode is similar to early simulations of dynamos driven by the
magneto-rotational instability \citep{BNST95}.

\section{Results}
\label{sect:results}

\subsection{Model without advective flux}

\begin{figure}
\includegraphics[width=\columnwidth]{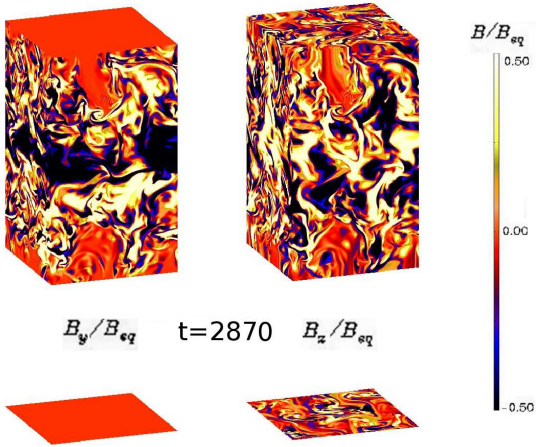}\\ 
\caption{Visualization of $B_y$ (left) and $B_z$ (right) on the 
borders of the domain for model N3 in the saturated phase of the 
simulation ($t$ is time in turnover times).
}\label{fig:bybz_dyn00a4}
\end{figure}

\begin{figure}
\includegraphics[width=\columnwidth]{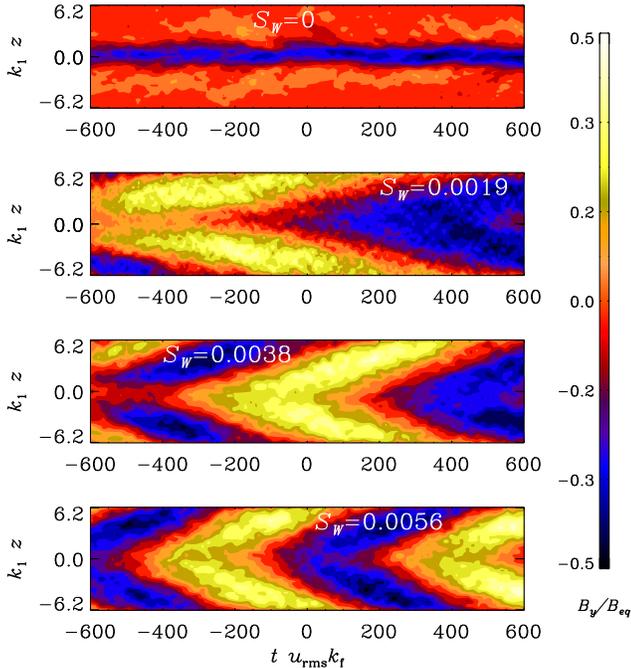}
\caption{Space-time diagrams of $\mean{B}_y$
for different wind intensities $\Shw$
corresponding to Models~N3, W3, M2, and S2 from top
to bottom.
The time axes have been shifted such that for each run about 1200
turnover times are being displayed.
Note that the cycle period decreases with increasing wind speed.
}\label{fig:pbut_multi}
\end{figure}

We begin by describing the results for
a dynamo in the absence of an advective flux ($\Shw=0$).
The solution for this particular setup is a steady
magnetic field mainly concentrated around the equator of the domain,
where the magnetic helicity changes its sign.
In \Fig{fig:bybz_dyn00a4} we show the $B_y$ and $B_z$
components of the magnetic field in the saturated phase 
of a model without wind and $\Rm=206$ (later referred to as Model~N3).
Note that $B_x=B_y=0$ on the top and bottom boundaries,
owing to the use of vertical-field boundary conditions.
Both of them, as well as $B_z$, do not show any significant 
temporal change once $\brms$ has reached its saturation value.
This can be observed in the top panel of \Fig{fig:pbut_multi},
where the vertical distribution of $\mean{B}_y$ is depicted as 
a function of time. 

The fact that this model is steady in the absence of a wind is surprising,
because according to linear mean-field calculations \citep{BCC09} it should
exhibit cyclic behavior with dynamo waves moving away from the midplane.
This discrepancy could be related to nonlinearity or to differences
resulting from the use of mean-field theory.
However, for the different boundary conditions used by MCCTB, mean-field
and direct numerical simulations exhibit rather similar behavior.
If it is a consequence of nonlinearity, it could be related to not
allowing magnetic helicity to escape the domain.
Indeed, the behavior is certainly quite different from the cases with
advective magnetic helicity flux (see below), and it is also
different from the otherwise similar accretion disc models.

\subsection{Model with advective flux}

\begin{figure}
\includegraphics[width=\columnwidth]{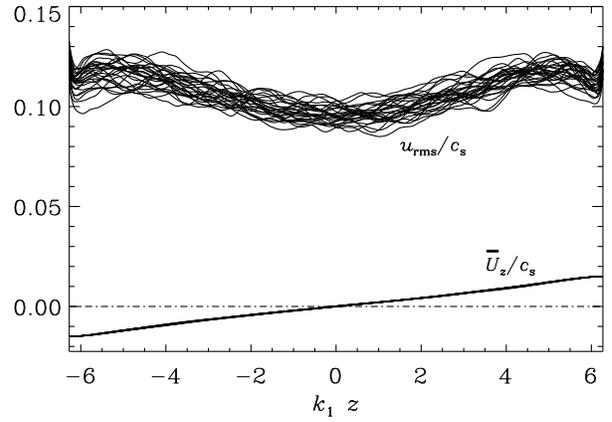}\\ 
\caption{Resulting vertical profile of $\mean{U}_z$
together with the rms velocity as a function of height.
Different lines correspond to different times.
In this case $U_0=0.015\cs$, corresponding to $\Shw=0.0055$.} 
\label{fig:pumean}
\end{figure}

Let us now turn to models in which a wind is included ($\Shw\neq0$).
An example of the resulting wind profile 
as well as the vertical distribution of $\urms$
is shown in \Fig{fig:pumean}.
Even with just a weak wind the dynamo becomes oscillatory;
see \Fig{fig:pbut_multi}.
Note that the cycle period decreases as the wind speed is increased.
We observe oscillatory solutions of even parity, that is 
$\mean{B}_x$ and $\mean{B}_y$ are on average symmetric with respect to
the midplane $z=0$, with dynamo waves migrating away from $z=0$.
This is expected based on mean-field models in similar setups \citep{BCC09}
provided the outer boundary condition is a vacuum or vertical field
condition, as is the case here.

\begin{figure*}
\includegraphics[width=2\columnwidth]{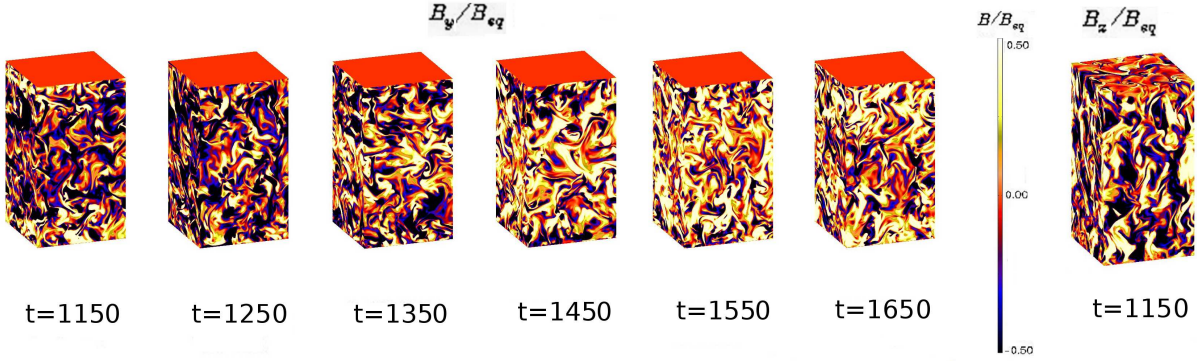}\\ 
\caption{
Visualization of $B_y$ (left) at six different times during the evolution of 
the system, for Model~S3. It is visible its variability, being this component
of the magnetic field prevalently negative in the first snapshot
and gradually turning positive in the others.
Time is given in turnover times and spans over half a cycle. 
On the right, $B_z$ is visualized on the 
borders of the domain for model S3.
It does not show any relevant variability during its evolution.
}\label{fig:bybz_dyn04b4_new}\end{figure*}

In \Fig{fig:bybz_dyn04b4_new} we can see how the actual $B_y(x,y,z,t)$,
as opposed to its horizontal average $\meanB_y(z,t)$, evolves during
half a period in 
the saturated phase of the simulations, changing gradually from negative
to positive polarity.
In \Tab{simulation} we summarize important output parameters that
characterize the simulations and, in particular, details regarding the
magnetic helicity balance.
Note that all table entries are non-dimensionalized by normalizing with
relevant quantities such as $\Beq$; see the table caption for details.
Magnetic helicity and the various production terms are antisymmetric about the
midplane.
Within the range $|z|\leq L_*$, all these quantities vary approximately
linearly with $z$.
Therefore we characterize their values by their slope.
An appropriate normalization is therefore $k_1\etatz\Beq^2$.

\begin{table*}\caption{
Characteristic output parameters of the simulations.
Here, $\Shw$ characterizes the wind speed,
$\meanBrms$ is the rms value of the {\it mean} field normalized by $\Beq$,
$2\memf\cdot\meanBB$, $2\eta\,\overline{\jj\cdot\bb}$, and
$\nab\cdot\meanFFFFf$ give magnetic helicity production,
dissipation, and flux divergence in units of $\kef\etatz\Beq^2$,
$\meanFFFdz$ is the turbulent--diffusive magnetic helicity flux
in units of $\etatz\Beq^2$,
characterized by the diffusion coefficient $\kappa_{\rm f}/\etat$,
$\keff$ is normalized by $\kef$,
$\overline{\jj\cdot\bb}$ is normalized by $\kef\Beq^2$,
and $\nab\cdot\meanFFFFm$ is the flux divergence of the mean
field in units of $\kef\etatz\Beq^2$.
The strongest outflows we simulate are those of Models~I1 and I2,
for which we reckon that the magnetic field is manly carried
out of the domain by the outflow.
The outflows on which we manly focus in this work are those of Models~S1 -- S6,
in which the maximum value of the wind speed is $U_0 \approx 0.15 \cdot \urms $.
$N_x$ indicates the number of mesh points in the $x$ direction.
(In all cases we have $N_y=N_x$ and $N_z=2N_x$.)
}\vspace{12pt}\centerline{\begin{tabular}{c|crccrrrlrc}
Model & $\Shw$ &  $\Rm$
& $2\memf\cdot\meanBB$ 
& $2\eta\,\overline{\jj\cdot\bb}$ 
& $\nab\cdot\meanFFFFf$
& $\meanFFFdz$
& $\kappa_{\rm f}/\etat$
& $\keff$
& $\overline{\jj\cdot\bb}$
& $\nab\cdot\meanFFFFm$\\
\hline
T1& 0.0000&$   9$&$ 0.066\pm0.019$&$-0.069\pm0.018$&$ 0.004\pm0.001$&$ 0.007$&$-0.3\pm0.7$&$ 1.22$&$-0.03$&$ 0.06$\\
T2& 0.0000&$  23$&$ 0.032\pm0.005$&$-0.035\pm0.003$&$ 0.004\pm0.007$&$ 0.002$&$-0.5\pm0.4$&$ 1.16$&$-0.03$&$ 0.03$\\
N1& 0.0000&$  37$&$ 0.048\pm0.007$&$-0.047\pm0.007$&$ 0.001\pm0.001$&$ 0.004$&$ 0.1\pm0.2$&$ 1.48$&$-0.07$&$ 0.05$\\
N2& 0.0000&$  81$&$ 0.023\pm0.008$&$-0.022\pm0.005$&$ 0.000\pm0.005$&$ 0.002$&$-0.0\pm0.2$&$ 1.37$&$-0.07$&$ 0.02$\\
N3& 0.0000&$ 206$&$ 0.001\pm0.001$&$-0.002\pm0.000$&$ 0.001\pm0.001$&$-0.003$&$ 0.0\pm0.2$&$0.97\ii$&$-0.01$&$0.00$\\
N4& 0.0000&$ 397$&$ 0.000\pm0.004$&$-0.001\pm0.001$&$-0.000\pm0.004$&$-0.000$&$ 0.1\pm0.2$&$0.84\ii$&$-0.01$&$0.00$\\
N5& 0.0000&$ 722$&$-0.006\pm0.002\ \ $\ &$-0.005\pm0.001$&$ 0.005\pm0.005$&$ 0.017$&$ 0.1\pm0.3$&$ 2.34$&$-0.16$&$-0.02\ $ \ \\
N6& 0.0054&$1073$&$ 0.010\pm0.004$&$-0.006\pm0.000$&$-0.018\pm0.015$&$ 0.019$&$ 0.0\pm0.3$&$ 2.83$&$-0.28$&$ 0.02$\\
W1& 0.0020&$  24$&$ 0.205\pm0.007$&$-0.196\pm0.007$&$-0.008\pm0.002$&$ 0.013$&$ 0.4\pm0.1$&$ 1.17$&$-0.19$&$ 0.19$\\
W2& 0.0019&$  51$&$ 0.094\pm0.022$&$-0.088\pm0.023$&$-0.005\pm0.001$&$ 0.011$&$ 0.5\pm0.1$&$ 1.45$&$-0.18$&$ 0.09$\\
W3& 0.0019&$ 129$&$ 0.047\pm0.004$&$-0.043\pm0.004$&$-0.004\pm0.002$&$ 0.010$&$ 0.4\pm0.2$&$ 1.60$&$-0.23$&$ 0.05$\\
W4& 0.0018&$ 265$&$ 0.026\pm0.002$&$-0.024\pm0.001$&$-0.003\pm0.003$&$ 0.008$&$ 0.2\pm0.2$&$ 2.03$&$-0.26$&$ 0.03$\\
W5& 0.0018&$ 540$&$ 0.014\pm0.004$&$-0.012\pm0.001$&$-0.002\pm0.012$&$ 0.008$&$ 0.0\pm0.2$&$ 2.62$&$-0.26$&$ 0.01$\\
M2& 0.0038&$  51$&$ 0.090\pm0.007$&$-0.082\pm0.010$&$-0.006\pm0.001$&$ 0.008$&$ 0.4\pm0.2$&$ 1.48$&$-0.17$&$ 0.09$\\
S1& 0.0060&$  24$&$ 0.167\pm0.004$&$-0.152\pm0.004$&$-0.012\pm0.002$&$ 0.019$&$ 0.8\pm0.2$&$ 1.27$&$-0.15$&$ 0.16$\\
S2& 0.0056&$  51$&$ 0.085\pm0.004$&$-0.074\pm0.004$&$-0.007\pm0.007$&$ 0.015$&$ 0.5\pm0.4$&$ 1.52$&$-0.16$&$ 0.08$\\
S3& 0.0055&$ 133$&$ 0.034\pm0.005$&$-0.029\pm0.004$&$-0.005\pm0.002$&$ 0.007$&$ 0.6\pm0.3$&$ 2.23$&$-0.16$&$ 0.03$\\
S4& 0.0053&$ 271$&$ 0.023\pm0.001$&$-0.018\pm0.001$&$-0.005\pm0.002$&$ 0.013$&$ 0.3\pm0.4$&$ 2.35$&$-0.20$&$ 0.02$\\
S5& 0.0053&$ 548$&$ 0.015\pm0.006$&$-0.011\pm0.000$&$-0.005\pm0.004$&$ 0.012$&$ 0.1\pm0.2$&$ 2.39$&$-0.25$&$ 0.02$\\
S6& 0.0054&$1063$&$ 0.013\pm0.003$&$-0.007\pm0.001$&$-0.006\pm0.009$&$ 0.010$&$ 0.1\pm0.2$&$ 2.70$&$-0.32$&$ 0.01$\\
I1& 0.0112&$  26$&$ 0.064\pm0.003$&$-0.060\pm0.002$&$-0.002\pm0.001$&$ 0.009$&$ 1.1\pm1.2$&$ 2.01$&$-0.06$&$ 0.06$\\
I2& 0.0105&$  55$&$ 0.029\pm0.007$&$-0.027\pm0.004$&$-0.002\pm0.004$&$ 0.007$&$-0.0\pm1.2$&$ 9.11$&$-0.06$&$ 0.03$\\
\label{simulation}\end{tabular}}\end{table*}

\begin{figure}
\includegraphics[width=\columnwidth]{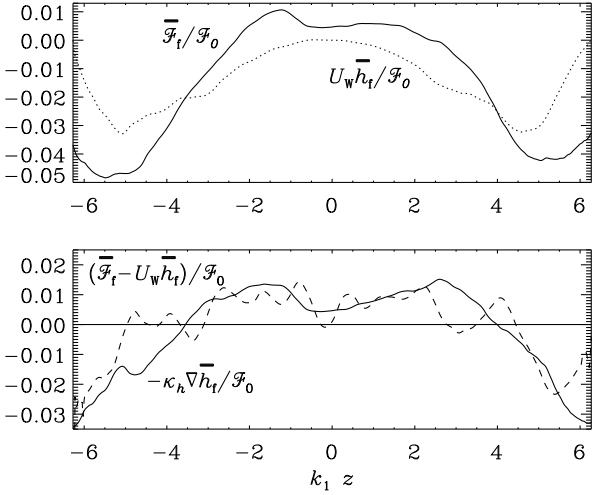}
\caption{
Contributions to the magnetic helicity flux for Model~W3.
Upper panel: vertical profiles of magnetic helicity fluxes
of the fluctuating field (solid line),
compared with the contribution from the mean flow (dashed line).
Lower panel: residual between the two aforementioned fluxes (solid line)
compared with a fit to the gradient of the magnetic helicity density
from the small-scale field (dashed line).
The fluxes are normalized by $\meanFFF_0=\etatz\Beq^2$.
}\label{fig:pflux_diff_dyn04c4}
\end{figure}

As can be seen from the bottom panel of \Fig{fig:pflux_diff_dyn04c4}, the
difference between the values of total and turbulent--diffusive fluxes
is roughly constant with $z$, so that its divergence is small.
This shows that in this particular setup the turbulent--diffusive
magnetic helicity flux has actually {\it no} contribution
in balancing the rhs of \Eq{balance} to zero.
This is different form the case studied by HB, in
which a finite magnetic helicity flux across the equator was possible,
playing thus a measurable role; see \Tab{Symmetry}.

To characterize the magnitude of the magnetic helicity, we give its value
averaged over the range $|z|\leq L_*$.
To compare this value with that from advective magnetic helicity fluxes,
we should multiply the table entry for $\nab\cdot\meanFFFFf$ by $L_*$,
which is about half the full vertical extent of the domain.
Note that $\nab\cdot\meanFFFFf$ and $k_1\meanFFFdz$ are actually
comparable, even though $\meanFFFdz$ can have no effect
in the present geometry and gives zero divergence.

We recall that $\overline{\jj\cdot\bb}$ and $\overline{\aaa\cdot\bb}$
are approximately proportional to each other.
This is also borne out by the present simulations where
$\keff^2\equiv\overline{\jj\cdot\bb}/\overline{\aaa\cdot\bb}$
is constant and $\keff/\kef\approx2$.
This confirms earlier findings of MCCTB and HB,
where a similar value of $\keff$ was found.
Under isotropic conditions, this ratio is approximately unity \citep{B01}.
However, for Models~N3 and N4, the correlation between $\overline{\jj\cdot\bb}$
and $\overline{\aaa\cdot\bb}$ is poor, giving formally a negative value,
so $\keff$ is given as imaginary in \Tab{simulation}.

The quantity $\overline{\jj\cdot\bb}/\kef\Beq^2$ is systematically
below unity, suggesting that the dynamo can only be expected to
produce mean fields where $\meanBB^2\approx\Beq^2$.
Finally, we also give the values of the flux divergence 
of the mean field $\nab\cdot\meanFFFFm$.
These values are typically about 10 times larger than the flux divergence
of magnetic helicity of the small-scale field, $\nab\cdot\meanFFFFf$,
but it is of course only the latter that is relevant 
for alleviating catastrophic quenching.

\begin{figure}
\includegraphics[width=\columnwidth]{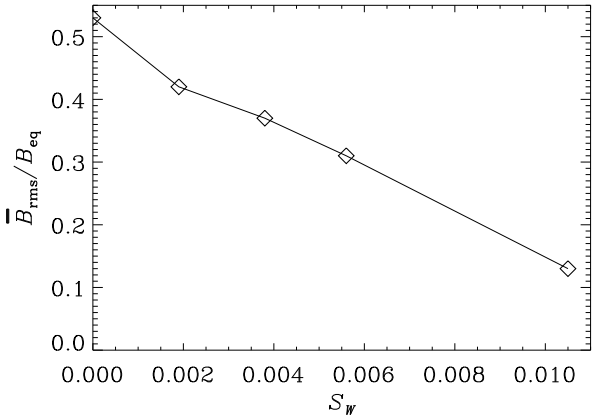}
\caption{
Root-mean-square value of the mean magnetic field, $\meanB_{\rm rms}$,
as function of $\Shw$
for models N3, W2, M2, S2 and I2, which have $\Rm \approx 50$.
}\label{fig:bvsunew}
\end{figure}

\begin{figure}
\includegraphics[width=\columnwidth]{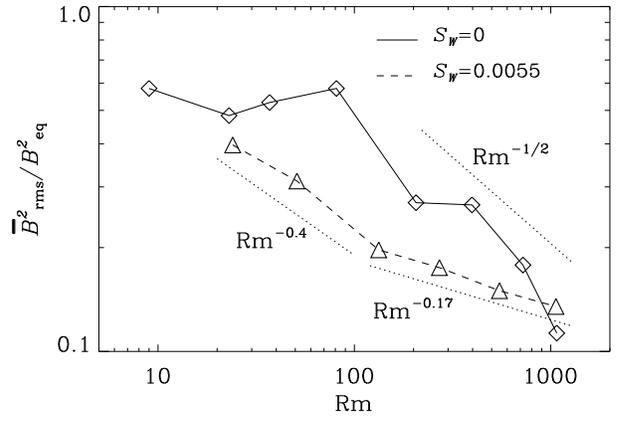}
\caption{
$\meanB_{\rm rms}^2$ as a
function of $\Rm$ in absence (solid line, Models~T1, T2, N1--N6)
and in presence (dashed line, Models~S1--S6) of advective flux.
The two dotted lines give the slopes $-0.5$, $-0.4$, and $-0.17$
for orientation.
}\label{fig:b_vs_rm}
\end{figure}

\begin{figure}
\includegraphics[width=\columnwidth]{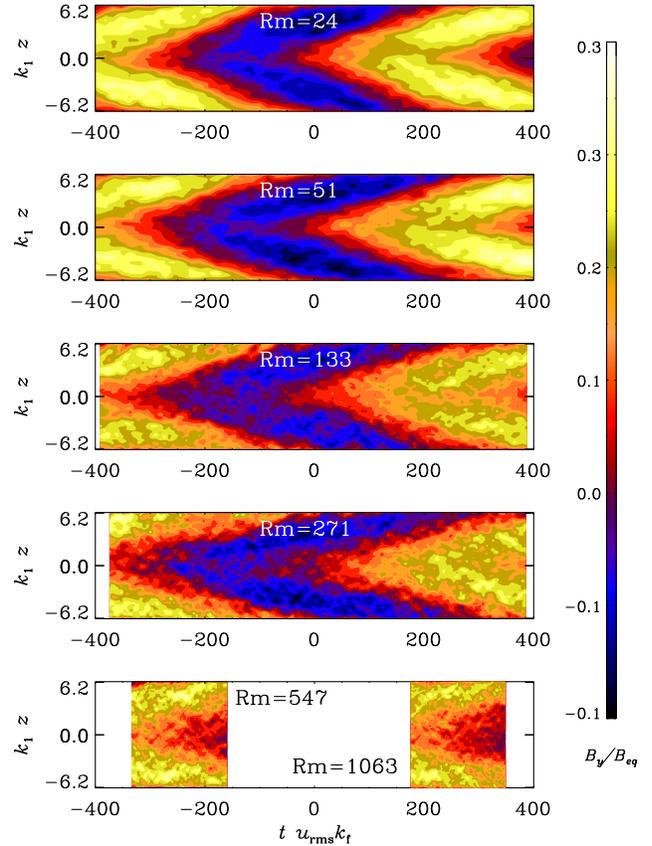}
\caption{Space-time diagrams of $\mean{B}_y$
for wind $U_0=0.015\cs$ (corresponding to $\Shw\approx0.0055$)
for Models~S1--S6
for different magnetic Reynolds numbers. From the top: $\Rm=24$, $51$, $133$,
271, as well as 547 (bottom left) and 1063 (bottom right).
}\label{fig:pbut_multi_rem}
\end{figure}

\begin{figure}
\includegraphics[width=\columnwidth]{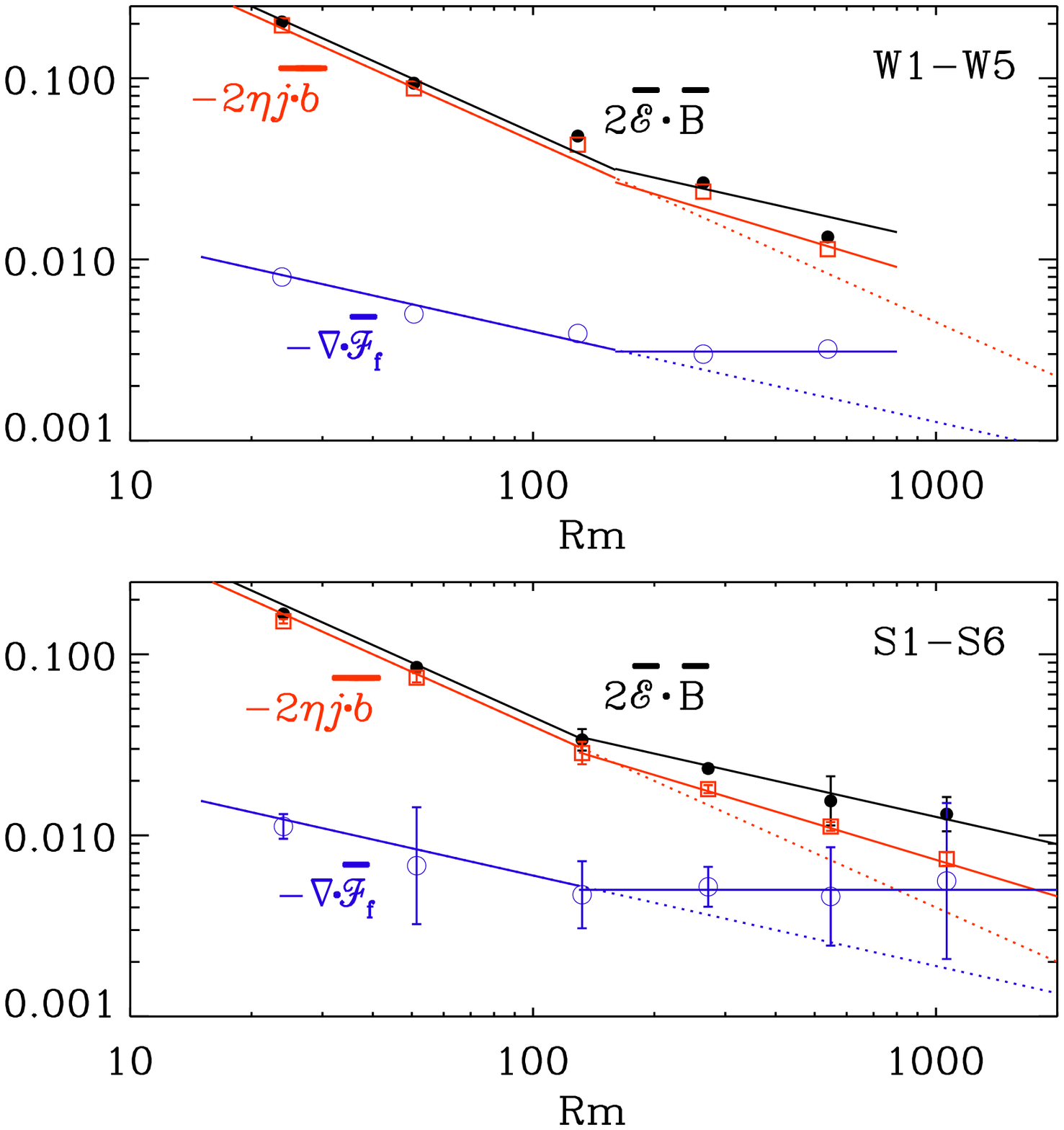}
\caption{
Scaling properties of the vertical slopes of $2\memf\cdot\meanBB$,
$-2\eta\mu_0\,\overline{\jj\cdot\bb}$, and $-\nab\cdot\meanFFFFf$
for Models~W1--W5 (upper panel)
and for Models~S1--S6 (lower panel).
(Given that the three quantities vary approximately linearly with $z$,
the three labels indicate their non-dimensional values at $k_1z=1$.)
The second panel shows that for a stronger wind
the contribution from the advective term becomes approximately
independent of $\Rm$ for $\Rm>170$ (blue line), while
that of the resistive term decreases approximately like $\Rm^{-2/3}$ (red line),
and $2\memf\cdot\meanBB$ decreases approximately like $\Rm^{-1/2}$ (black line).
}\label{fig:presults}
\end{figure}

All simulations with wind show that the rms value of the mean field,
$\meanBrms$, declines slowly with increasing wind speed;
see \Fig{fig:bvsunew}.
This result might just be a consequence of a gradual increase of the critical
value of $\Rm$ above which dynamo action is possible.
However, it could also be an indication that a fraction of the mean 
magnetic field is being removed from the domain by the flow -- as
found in the mean-field models of \cite{SSSB06}.

In \Fig{fig:b_vs_rm} we see how $\meanBrms$ decreases with increasing $\Rm$.
The scalings $\Rm^{-0.4}$ and $\Rm^{-0.17}$ are given for orientation and
show that in the presence of advection $\meanBrms$ varies much slower than
$\Rm^{-1}$, which is the slope anticipated from catastrophic quenching
models without a wind \citep{BS05c}.
Note, however, that DNS always gave a shallower slope \citep{BD01} and,
at larger values of $\Rm$, $\meanBrms$ may have been already independent
of $\Rm$ \citep{HB12}.
Indeed, without a wind ($\Shw=0$) the $\Rm$ dependence is compatible with
a steeper $\Rm^{-1/2}$ law, but it is less certain in this case.
Looking at \Fig{fig:pbut_multi_rem}, we can also see that there is
no significant change of the cycle period with $\Rm$.
The high-resolution runs with $\Rm=544$ and 1061 are too short to cover
a magnetic cycle, but one can see that the slope of the structure, which
corresponds to the speed of the dynamo wave, is approximately unchanged.
In the high-$\Rm$ models the fluctuations are more pronounced, but the
peak-to-peak contrast is about the same for all runs.

\Tab{simulation} shows that $2\memf\cdot\meanBB$,
$2\eta\mu_0\,\overline{\jj\cdot\bb}$, and $\nab\cdot\meanFFFFf$
balance approximately to zero, confirming that the results represent
a statistically steady state.
All three quantities have approximately the same (nearly linear) $z$ dependence
for $|z|<L_*$, so that also the values of their three slopes must balance
to zero, which is indeed the case.
In \Fig{fig:presults} we show the scaling properties of the
aforementioned quantities for Models~W1--W5 and S1--S6.
For $\Rm\lesssim\Rmast$, where $\Rmast\approx170$ for $\Shw=0.002$
and $\Rmast\approx120$ for $\Shw=0.005$,
the first two quantities decrease approximatively
like $\Rm^{-1}$, while the latter decreases only like $\Rm^{-1/2}$,
which is in agreement with the values obtained by HB; see also
Figure~10 of \cite{CHBK11} for a corresponding plot.

\begin{table}\caption{
Additional parameters of the simulations including $\Rm$,
magnetic diffusivity, the ratios `$\meanB/u$' ($=\meanBrms/\urms$)
and `$\meanB/b$' ($=\meanBrms/\brms$), as well as
Mach number and number of mesh points.
}\vspace{12pt}\centerline{\begin{tabular}{c|rccccr}
Model & $\Rm$ & $\eta k_1/\cs$ & `$\meanB/u$' & `$\meanB/b$' & $\Ma$ & $N_x$ \\
\hline
T1&$   9$&$5.0\times10^{-3}$&$0.58$&$1.13$&$ 0.18$&$  64$\\
T2&$  23$&$2.0\times10^{-3}$&$0.48$&$0.87$&$ 0.19$&$  64$\\
N1&$  37$&$1.0\times10^{-3}$&$0.53$&$0.70$&$ 0.15$&$  64$\\
N2&$  81$&$5.0\times10^{-4}$&$0.58$&$0.73$&$ 0.16$&$ 128$\\
N3&$ 206$&$2.0\times10^{-4}$&$0.27$&$0.33$&$ 0.17$&$ 256$\\
N4&$ 397$&$1.0\times10^{-4}$&$0.27$&$0.33$&$ 0.16$&$ 512$\\
N5&$ 722$&$5.0\times10^{-5}$&$0.18$&$0.21$&$ 0.15$&$1024$\\
N6&$1073$&$2.5\times10^{-5}$&$0.11$&$0.15$&$ 0.11$&$1024$\\
W1&$  24$&$1.0\times10^{-3}$&$0.61$&$0.69$&$ 0.10$&$ 128$\\
W2&$  51$&$5.0\times10^{-4}$&$0.42$&$0.48$&$ 0.10$&$ 128$\\
W3&$ 129$&$2.0\times10^{-4}$&$0.36$&$0.39$&$ 0.10$&$ 256$\\
W4&$ 265$&$1.0\times10^{-4}$&$0.28$&$0.31$&$ 0.11$&$ 512$\\
W5&$ 540$&$5.0\times10^{-5}$&$0.19$&$0.22$&$ 0.11$&$1024$\\
M2&$  51$&$5.0\times10^{-4}$&$0.36$&$0.45$&$ 0.10$&$ 128$\\
S1&$  24$&$1.0\times10^{-3}$&$0.40$&$0.55$&$ 0.10$&$  64$\\
S2&$  51$&$5.0\times10^{-4}$&$0.31$&$0.42$&$ 0.10$&$ 128$\\
S3&$ 133$&$2.0\times10^{-4}$&$0.20$&$0.27$&$ 0.11$&$ 256$\\
S4&$ 271$&$1.0\times10^{-4}$&$0.17$&$0.23$&$ 0.11$&$ 512$\\
S5&$ 548$&$5.0\times10^{-5}$&$0.15$&$0.19$&$ 0.11$&$1024$\\
S6&$1063$&$2.5\times10^{-5}$&$0.14$&$0.17$&$ 0.11$&$1024$\\
I1&$  26$&$1.0\times10^{-3}$&$0.18$&$0.36$&$ 0.10$&$  64$\\
I2&$  55$&$5.0\times10^{-4}$&$0.13$&$0.26$&$ 0.11$&$ 128$\\
\label{simulation2}\end{tabular}}\end{table}

However, for $\Rm\gtrsim 200$ the scaling of $2\memf\cdot\meanBB$ changes
into an $\Rm^{-1/2}$ scaling; $\nab\cdot\meanFFFFf$ is at first
below $2\eta\mu_0\,\overline{\jj\cdot\bb}$, but for high enough $\Rm$
increases to reach an absolute value similar to that of $2\memf\cdot\meanBB$.
This suggests that the simple expectation based on the naive extrapolation
given from a linear fit is misleading, and that catastrophic quenching
might be alleviated already for $\Rm\gtrsim 1000$.
In the absence of a wind and for large magnetic Reynolds numbers
(Models~N4--N6), the divergence of the magnetic helicity flux shows
strong fluctuations about zero, making it harder to determine an
accurate magnetic helicity balance of small-scale fields.

In \Tab{simulation2} we summarize additional output parameters
of the simulations including $\Rm$, the magnetic diffusivity,
the ratios of the rms values of mean field to fluctuating velocity
and fluctuating magnetic field, i.e., $\meanBrms/\urms$ and
$\meanBrms/\brms$, respectively, as well as Mach number and
number of mesh points.
As was already obvious from \Fig{fig:b_vs_rm}, $\meanBrms/\urms$
(which is the same as $\meanBrms/\Beq$), decreases with increasing
$\Rm$, and the same is also true of the ratio $\meanBrms/\brms$.
The numerical resolution in the $x$ direction, $N_x$, is given in the
last column.
This is also the resolution used in the $y$ direction, while that in the
$z$ direction is always twice as large.

\section{Conclusions}
\label{sect:concl}

In the present work we have examined the effects of an advective
magnetic helicity flux in DNS of a turbulent dynamo.
The present simulations without shear yield an oscillatory
large-scale field owing to the spatially varying kinetic
helicity profile with respect to the equatorial plane.
We emphasize in this context that the possibility of oscillatory
dynamos of $\alpha^2$ type is not new
\citep{Baryshnikova+Shukurov87,Raedler+Brauer87}, but until recently
all known examples were restricted to spherical shell dynamos
where $\alpha$ changes sign in the radial direction.
The example found by \cite{Mitra+etal_10b} applies to a spherical wedge 
with latitudinal variation of $\alpha$ changing sign about the equator.
Similar results have also been obtained in a mean-field dynamo
with a linear variation of $\alpha(z)\propto z$ \citep{BCC09}.
Our present simulations are probably the first DNS of such a dynamo
in Cartesian geometry.
Closest to our simulations are those of MCCTB who used
perfectly conducting outer boundary conditions without wind, and also
found oscillatory solutions.
Surprisingly, however, oscillations are here only obtained
if there is at least a slight outflow.

One would have expected that catastrophic quenching can be alleviated if
magnetic helicity is removed from the domain at a rate
larger than its diffusion rate, that is, the advective term
$\nab\cdot\meanFFFFf$ dominates over the resistive term,
$2\eta\mu_0\,\overline{\jj\cdot\bb}$.
\Fig{fig:presults} shows that, for $\Rm\lesssim 200$, the latter term
decreases linearly with decreasing $\eta$, while the former only decreases
proportional to $\eta^{1/2}$, i.e., proportional to $\Rm^{-1/2}$.
This would have led us to the estimate that for $\Rm\approx 4\cdot10^3$ the
catastrophic quenching can be alleviated by a wind with $\Shw\approx 0.005$.
Our new results suggest that this can happen
already for smaller values of $\Rm$.
The reason for this is still unclear.
It is possible that catastrophic quenching was an artefact of intermediate
values of $\Rm$, as suggested by \cite{HB12}, or that a magnetic helicity
flux can have an effect even though it is weak compared with diffusive terms.

Finally, we should emphasize that we have only examined here the case
of subsonic advection.
In real astrophysical cases, like galactic and stellar winds,
the outflow is instead supersonic and can, thus, play 
an even more important role in alleviating the catastrophic quenching
through the advection of magnetic helicity.
This assumes, of course, that the dynamo is strong enough to be still
excited in the presence of a stronger wind.

\section*{Acknowledgements}

FDS acknowledges HPC-EUROPA for financial support.
Financial support from European Research Council under the AstroDyn
Research Project 227952 is gratefully acknowledged.
The computations have been carried out at the National Supercomputer
Centre in Ume{\aa} and at the Center for Parallel Computers at the
Royal Institute of Technology in Sweden.

\def\apj{ApJ}
\def\apjl{ApJL}
\def\pre{Phys. Rev. E}
\def\aap{A\&A}
\def\mnras{MNRAS}
\def\solphys{Sol. Phys.}
\bibliographystyle{mn2e}

\label{lastpage}
\end{document}